# A framework to discover potential ideas of new product development from crowdsourcing application

Thanh-Cong Dinh[1], Hyerim Bae[1], Jaehun Park[1] and Joonsoo Bae[2]

[1] *Department of Industrial Engineering, Pusan National University, South Korea*
E-mail: cong.dinh@pusan.ac.kr, hrbae@pusan.ac.kr, pjh3479@pusan.ac.kr

[2] *Department of Industrial and Information Systems Engineering, Chonbuk National University, South Korea*
E-mail: jsbae@chonbuk.ac.kr

**Abstract**

In this paper, we study idea mining from crowdsourcing applications which encourage a group of people, who are usually undefined and very large sized, to generate ideas for new product development (NPD). In order to isolate the relatively small number of potential ones among ideas from crowd, decision makers not only have to identify the key textual information representing the ideas, but they also need to consider online opinions of people who gave comments and votes on the ideas. Due to the extremely large size of text data generated by people on the Internet, identifying textual information has been carried out in manual ways, and has been considered very time consuming and costly. To overcome the ineffectiveness, this paper introduces a novel framework that can help decision makers discover ideas having the potential to be used in an NPD process. To achieve this, a semi-automatic text mining technique that retrieves useful text patterns from ideas posted on crowdsourcing application is proposed. Then, we provide an online learning algorithm to evaluate whether the idea is potential or not. Finally to verify the effectiveness of our algorithm, we conducted experiments on the data, which are collected from an existing crowd sourcing website.

## 1. Introduction

The term "crowdsourcing" was firstly introduced by Jeff Howe to indicate the act of outsourcing a task traditionally performed by an employee or contractor to a large undefined group of people (i.e., a crowd) [1]. Crowdsourcing applications (e.g. Dell's IdeaStorm [2], Starbuck's MyStarbucksIdea [3], Quirky, etc.) encourage individuals to suggest ideas for new product development (NPD), by using online interface. Though crowdsourcing has widely been used for collecting variety of ideas from many people, there are drawbacks of participants' lack of expertise in the subject area and correspondingly poor quality of solutions. Therefore, before considering new and interesting ideas from crowd, firms have to carefully screen them. Unfortunately, the great mass of unstructured data (e.g. textual information) generated by such applications is very difficult to screen or evaluate, and this is not a simple problem of admittance to the data. This difficulty is reflected in the relative paucity of case studies on, for example, Dell's IdeaStorm [4-6]. Some previous studies on technical idea mining [7, 8] and new product idea screening [9, 10] are at very early stage not considering the crowdsourcing environment, and thus idea mining from crowd could not properly processed.

The motivation of the present study is to formulate a new method to discover potential NPD ideas in crowdsourcing applications. For this purpose, we developed a novel, semi-automatic text-mining- and prediction-model-based approach to save decision makers' time in screening masses of ideas and help them determine which ideas are potentially implementable for NPD. The followings are three main contributions of this paper.

- First, we introduce a text mining method combined with an existing computational linguistics techniques to extract useful information from online crowd sourcing applications.
- Second, a set of measurements is devised and formulated for evaluating ideas, which can be classified into two types. The first type is relevance measurement, which is used to measure how the ideas are suitable for the firm. The second type is interest measurement, which is used to measure the degree of crowd's interest on the idea.
- Third, we developed an online learning algorithm by extending logistic regression model to calculate the probability of an idea's being potential based on the measurements formulated in this paper.

The rest of this paper is organized as follows: Section 2 overviews the related work, Section 3 introduces our methodology, and Section 4 discusses the pertinent experimentation and results; lastly, Section 5 draws conclusions and anticipates future work.

## 2. Related work

## 2.1 Idea mining and text mining

Thorleuchter *et al.* [50 – 53] introduced the concept of "idea mining" which is an automatic process of extracting new and useful ideas from unstructured text. They defined idea as a combination of two attributes: a mean and an appertaining purpose [50]. The idea mining process contains three steps: the preparation of a problem description, the extraction of text patterns from a new text and the evaluation of text patterns for novelty and usefulness concerning problem description. Their approach can be simplified as a text mining process of finding dissimilarity of text terms between a problem description and a problem solution idea.

Tseng *et al.* [54] utilized text mining techniques to analyze patent documents. Their approach includes text segmentation, summary extraction, feature selection, term association, cluster generation, topic identification, and information mapping. In the same manner, Lee *et al.* [55] proposed a text mining approach for developing keyword-based patent maps for use in new technology creation activity.

## 2.2 Behavior of crowdsourcing

Di Gangi and Wasko [33] initially utilized data collected from the website IdeaStorm of Dell to identify the factors that influence an organization's adoption decision when innovations come from outside of the organization. In another context, Bayus [34, 35] studied past success behavior of the ideators who posted ideas on Dell's IdeaStorm website. The empirical results show that ideators who have past success end up with ideas similar to their previous ideas (i.e. they generate less diverse ideas). However, this negative effect of past success is somewhat mitigated for ideators who comment on other ideas.

In Huang *et al.* [36], the authors proposed that the firms should reduce their response time to the ideas, and thus it would lead to more contributions of the ideas to the firm. As a summarization of this section, a comparison between this paper and relevant researches is given in Table 1. The columns indicate the problems, which the authors considered, and the rows express main methodologies they dealt with.

Table 1. Comparison between this paper and relevant researches

| Methodology | Ideas for NPD | | | Ideas for problem solving | Patent analysis |
| --- | --- | --- | --- | --- | --- |
| | From crowdsourcing | From internal | From customers | | |
| Text mining | This paper | | [52, 53] | [50, 51] | [54, 55] |
| Relevance measurement | This paper | | | | |
| Interest measurement | [34] This paper | | | | |
| Online logistic regression | This paper | | | | |
| Knowledge Base | | [40] | | | |
| Markov model | | [43, 44] | | | |
| Fuzzy group-preferences | | [42] | | | |

## 3. Idea mining from crowd applications

In order to extract ideas from an online crowd application, we develop a three-step procedure, which is schematized in Fig. 1.

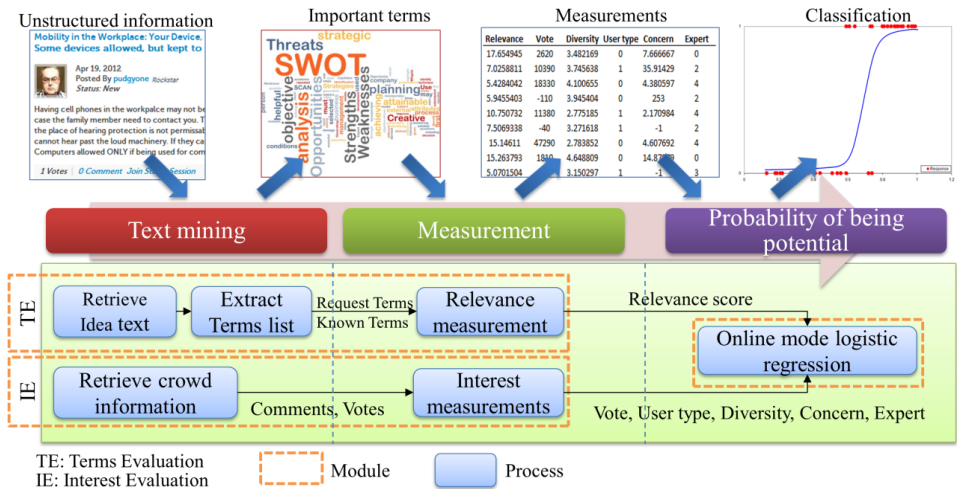

Fig. 1. Framework to discover potential new product ideas from crowdsourcing application

The first step is text mining, which extracts important terms from textual data. The second step is evaluating terms by numerical method based on the measurements developed in this paper. The purpose of these two steps is to transform textual information into a set of measurements. To this end, the following two main modules are proposed:

- Terms Evaluation (TE): the goal of this module is to calculate the relevance score of important terms. To achieve this goal, first, title body of a target idea in a textual format is retrieved from the web page. It is then examined to search for Request Terms (RTs), which are the request phrases employed by users. Known Terms (KTs) are also extracted from the idea text, which are phrases familiar to the firm.
- Interest Evaluation (IE): this module is to measure the interest of crowd on the idea, it includes two processes: retrieve crowd information and interest measurement. The information from the crowd (e.g. comments, votes) of a target idea is retrieved from its web page. Then, the number of votes and comment activities (e.g. comments, the profile of commenters, etc.) are collected for measurement of interest of the crowd on the target idea.

Finally, an online mode logistic regression (OLR) algorithm is applied to calculate the probability of being potential in consideration of both relevance and interest measurements.

### 3.1 Terms evaluation

#### 3.1.1 Retrieve idea text

This section explains two core definitions in Definition 1 and 2, which are used to present the unstructured textual information in a structured way.

**Definition 1 (Idea text)**

An idea text $I = \{s_1, \ldots, s_n\}$ is a list of sentences $(s_i)$ in order of appearance and let $n$ be the number of sentences in $I$ and $i = 1, \ldots, n$. The first sentence $s_1$ is usually used as the idea title and the others $(s_i, \forall i > 1)$ are used for expressing the idea body.

Fig. 2 shows an example (Ex1) of idea text in which the idea title is mapped to $s_1$, and other six sentences in an idea body are mapped to $s_2$ through $s_7$. Notice that we have: $I = \{s_1, s_2, \ldots, s_7\}$ = {"Sell the first anti-microbial keyboard", "Dell",... , "This will get Dell in the medical journals as being cutting edge in health well being"}

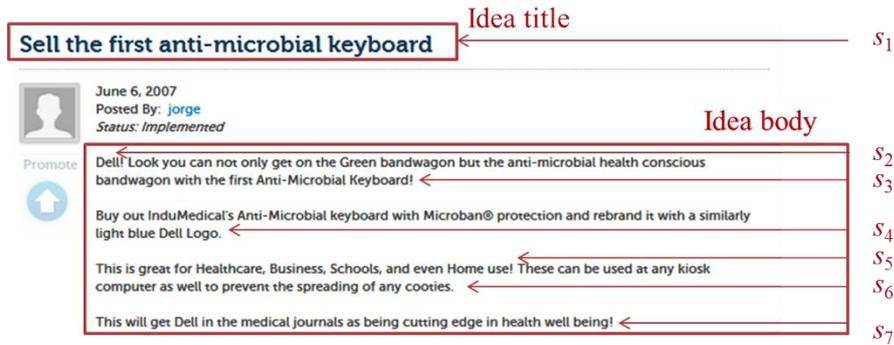

Fig. 2. An example of idea text (Ex1)

**Definition 2 (Sentence)**

A sentence is defined as a list of tokens $(t_i)$, that is, $s = \{t_1, \ldots, t_m\}$, where $m$ is the number of tokens in $s$ and $i = 1, \ldots, m$. An arbitrary token $t_i$ is expressed by a tuple $(w, pt)$, where
(i) $w$ is the word that the token represents.
(ii) $pt$ is the Penn Treebank tag [56] of $w$.
We define two functions
(i) $word(t_i)$: a function to retrieve $w$ of $t_i$
(ii) $tag(t_i)$: a function returning $pt$ of $t_i$
In this paper, $t_i$ is sometimes used for simple notation for $word(t_i)$.

From Ex1 in Fig. 2, the sentence $s_1$ ("Sell the first anti-microbial keyboard") is a set with two functions. That is, $s_1 = \{t_1, t_2, \ldots, t_5\} = $ {("sell", VB), ("the", DT), ("first", JJ), ("anti-microbial", JJ), ("keyboard", NN)}. Functions $word(t_4)$ and $tag(t_4)$ return "anti-microbial" and JJ, respectively. According to [56], the tag JJ expresses adjective.

**3.1.2 Extract term list**

This section describes core definitions, which are used to extract important terms (or keywords) from structured data, and two text mining algorithms are also proposed in the section. Definition 3 is a definition for terms that are extracted from a sentence.

**Definition 3 (Term)**

A term $T$ is defined as a phrase of several tokens without their tagging with certain types of tag. Every token $t_i$ in a term should satisfy $tag(t_i) \in (J \cup R \cup N \cup C)$, $J$ is a set of adjective tags (i.e. JJ, JJR, JJS); $R$ is a set of adverb tags (i.e. RB, RBR, RBS); $N$ is a set of noun tags (i.e. NN, NNS, NNP, NNPS) and $C$ is a set of cardinal number and possessive ending tags (i.e. CD, POS).

In Ex 1, only one term can be extracted, that is "anti-microbial keyboard".

**Definition 4 (Characteristic adjectives/ adverbs list and stop-words list)**

**Characteristic adjectives/ adverbs list ($E$):** $E$ is a pre-defined dictionary which is a list of characteristic adjective and adverb words that often occur together with requests from customers (i.e. thinner, lighter, longer, cheaper etc.).

**Stop-words list ($S$):** $S$ is a list of stop-words which are words that should be filtered out from the text (i.e. "a", "about", "above", "all", "an", etc.).

In order to calculate the value of relevant measurement, we need more definitions about Known Term, Request Term, Terms list, and Word frequency, which are provide in Definition 5, 6, 7, and 8 respectively.

**Definition 5 (Known Term Dictionary and Known Term)**

**Known Term Dictionary** (*K*): *K* is a pre-defined dictionary, which is a list of commonly used words (or terms) by the firm and each token in a term has a weight value returned by using a function $w(t_i)$. *K* usually represents current technologies as well as current products of the firm.

**Known Term** (*KT*): *KT* is a special term $KT = \{t_i | i=1,...,J\}$ which satisfies the following conditions: There exists a token $t_i$, $i \in [1,...,J]$ such that $t_i \in K$, and there does not exist any one token $t_j$, $j \in [1,...,J]$ such that $tag(t_j) \in J \cup R$. That means a *KT* is composed of a set of tokens in which at least one token is belong to a dictionary *K*; and there is no token which is either adjective or adverb or possessive.

In Ex1,, the term "keyboard" is an element of *KT*.

**Definition 6 (Request Term)**

A Request Term (*RT*) is a term which expresses request from an individual. The request can be about adding an optional technology, or offering a new accessory, etc.

In order to find *RT*, we first detect characteristic words such as preposition or subordinating conjunction words (e.g. "into", "to", "for") or verbs (e.g. "need", "offer") in the sentence. Then, perform searching for noun phrases around those words. In Ex1, the term "anti-microbial keyboard" is an element of *RT*.

**Definition 7 (Terms list)**

A terms list is a list of criteria attributes. In our paper, terms list is represented as *V(α, β)*, where *α* is a list of distinct Request Terms and *β* is a list of distinct Known Terms.

In Ex1, *V(α, β)* = ({"anti-microbial keyboard"}, {"keyboard"})

**Definition 8 (Word frequency)**

Word frequency function $wf(t_i)$ returns the number of occurrence of the word $t_i$ in the whole text *I*.

In Ex1, *wf*("anti-microbial") = 4 and *wf*("keyboard") = 3.

To extract terms list, we use two algorithms. The first algorithm named *ExtractIdeaTitle* is used when the number of tokens in an idea title is greater than or equal to two. In other cases, the second algorithm *ExtractWholeIdea* is employed.

For the first algorithm, two functions are used to create request terms and known terms. Function 1 (*createRequestTerm*) is used to create request terms from an arbitrary string which was already tokenized and applied tags. According to Definition 6, the function should retrieve as many meaningful noun phrases as possible. Hence, Function 1 tries to create noun phrases which are combined by not only nouns but also adverbs, adjectives and etc. WordNet® database [57] was employed to find the base form of the adjectives (e.g. base form of the word "thinner" is "thin"). Moreover, as in Definition 4, only some characteristic adjectives are used to create the terms.

---

**Function 1** createRequestTerms(*t*, *start*, *end*)

    **Input**: *t* is a list of tokens, the *start* index and *end* index in *t*
    **Output**: terms list *RT*

1    $RT \leftarrow \{\varnothing\}$ /* Create an empty list of request terms string*/
2    $JJ \leftarrow \{\varnothing\}$ /* Create an empty list of adjectives or adverbs string*/
3    $PosJJ \leftarrow \{\varnothing\}$ /* Create an empty list of indexes which are associated with *JJ* */
4    $NN \leftarrow \{\varnothing\}$ /* Create an empty list of nouns string */
5    $PosNN \leftarrow \{\varnothing\}$ /* Create an empty list of indexes which are associated with *NN* */
6    $tmp \leftarrow \{\varnothing\}$ /* Create an empty temporary string */
7    **FOR** $i \leftarrow start$ **TO** *end* **DO**

| | |
|---|---|
| 8 | **CASE** $e \leftarrow tag(t_i)$ **OF** /* Definition 3 */ |
| 9 |   $e \in J$ : |
| 10 |     $t_i \leftarrow$ getBaseForm$(t_i)$ /* get base form of $t_i$ using WordNet® database [57] */ |
| 11 |     **IF** $t_i \in E$ **THEN** |
| 12 |       $JJ \leftarrow t_i$ /* Definition 4 */ |
| 13 |       $PosJJ \leftarrow i$ |
| 14 |     **ENDIF** |
| 15 |   $e \in R \cup N \cup C$ : |
| 16 |     **IF** $tmp = \{\varnothing\}$ **THEN** $PosNN \leftarrow i$ |
| 17 |     $tmp \leftarrow t_i$ |
| 18 |   **OTHER** : |
| 19 |     **IF** $tmp \notin S$ **THEN** $NN \leftarrow tmp$ /* Definition 4 */ |
| 20 |     $tmp \leftarrow \{\varnothing\}$ |
| 21 |   **ENDCASE** |
| 22 |   **IF** $i \geq end$ **THEN** |
| 23 |     **IF** $tmp \notin S$ **THEN** $NN \leftarrow tmp$ |
| 24 |     $tmp \leftarrow \{\varnothing\}$ |
| 25 |   **ENDIF** |
| 26 | **ENDFOR** |
| | /* Create noun phrases as mentioned in Definition 6 */ |
| 27 | **FOR** $i \leftarrow 1$ **TO** $|NN|$ **DO** /* $|X|$ returns the cardinality of $X$ */ |
| 28 |   **IF** $|JJ| > 0$ **THEN** |
| 29 |     **FOR** $j \leftarrow 1$ **TO** $|JJ|$ **DO** |
| 30 |       **IF** $PosJJ_j < PosNN_i$ **THEN** $RT \leftarrow$ concatnate$(JJ_j, NN_i)$ /* concatenate two strings with a space character between them /* |
| 31 |     **ENDFOR** |
| 32 |   **ELSE** |
| 33 |     $RT \leftarrow NN_i$ |
| 34 |   **ENDIF** |
| 35 | **ENDFOR** |
| 36 | **RETURN** $RT$ |

Function 2 is used to create known terms from a string. This string is already tokenized and applied tags.

**Function 2** createKnownTerms($t$, $start$, $end$)

**Input**: $t$ is a list of tokens, the *start* index and *end* index in $t$
**Output**: terms list $KT$

| | |
|---|---|
| 1 | $KT \leftarrow \{\varnothing\}$ /* Create an empty list of known terms string*/ |
| 2 | $tmp \leftarrow \{\varnothing\}$ |
| 3 | **FOR** $i \leftarrow start$ **TO** $end$ **DO** |
| 4 |   **IF** $t_i \in U$ **THEN** /* Definition 3 */ |
| 5 |     **IF** $t_i \in N \cup C$ **THEN** $tmp \leftarrow t_i$ /* Definition 5 */ |
| 6 |   **ELSE** |
| 7 |     **IF** $tmp \notin S$ **THEN** $KT \leftarrow tmp$ /* Definition 4 */ |
| 8 |     $tmp \leftarrow \{\varnothing\}$ |
| 9 |   **ENDIF** |
| 10 |   **IF** $i \geq end$ **THEN** |
| 11 |     **IF** $tmp \notin S$ **THEN** $KT \leftarrow tmp$ |
| 12 |     $tmp \leftarrow \{\varnothing\}$ |
| 13 |   **ENDIF** |
| 14 | **ENDFOR** |
| 15 | **RETURN** $KT$ |

Using Function 1 and Function 2, Algorithm 1 can be presented as follows

**Algorithm 1** ExtractIdeaTitle($I$)

    **Input**: an idea text $I$, $I = \{s_1,\ldots,s_n\}$ /* Definition 1 */
    **Output**: terms list $V$

1    $KT \leftarrow \{\varnothing\}$
2    $RT \leftarrow \{\varnothing\}$
3    $str \leftarrow \text{cleaning}(s_1)$ /* Remove all non-alphabetic characters except space, dot, hyphen and apostrophe from idea title $s_1$ */
4    $t \leftarrow \text{tokenize}(str)$ /* Tokenize $str$ into single words and store into array $t$ */
5    $t \leftarrow \text{pennTag}(t)$ /* Apply Penn Treebank tags on $t$ using the algorithm from [58] */
6    **IF** $\exists c, c \in \{\text{"for"}, \text{"into"}\}: c \in t$ **THEN** /* Case 1*/
7       $flag \leftarrow \text{position}(c)$ /* Get position of $c$ in $t$ */
8       $RT \leftarrow \text{createRequestTerms}(t, 1, flag - 1)$
9       $KT \leftarrow \text{createKnownTerms}(t, flag + 1, |t|)$
10  **ELSE IF** $\exists c, c \in \{\text{"need"}, \text{"offer"}\}: c \in t$ **THEN** /* Case 2 */
11    $flag \leftarrow \text{position}(c)$
12    $RT \leftarrow \text{createRequestTerms}(t, flag + 1, |t|)$
13    $KT \leftarrow \text{createKnownTerms}(t, 1, flag - 1)$
14  **ELSE** /* General case */
15    $RT \leftarrow \text{createRequestTerms}(t, 1, |t|)$ /* Create temporary terms list */
16    **FOR** each $term \in tmp$ **DO**
17      **IF** $term \in K$ **THEN** $KT \leftarrow term$ /* Definition 5 */
18      **ELSE** $RT \leftarrow term$
19      **ENDIF**
20    **ENDFOR**
21  **ENDIF**
22  **RETURN** $V \leftarrow RT, KT$

To illustrate how algorithm1 works, three idea titles "Touchscreen Option For Dell Inspiron Mini / Dell E / Dell E Slim.", "Dell should offer the XPS m1330 with a 14.1 inch screen", and "Sell the first anti-microbial keyboard" are considered. After applying tokenization and tagging, the procedure of creating terms list for each idea title are illustrated in Fig 3(a), (b) and (c), respectively. When an idea has very short title or is not in common format, we cannot use Algorithm 1. Algorithm 2 is employed for supporting such cases. The idea of Algorithm 2 is that we firstly extract $k$ keywords, each of which is a single-word that has a high value of word frequency $wf$ (e.g. $wf \geq 2$) in the body of idea text. Then, the position of the keyword in the text is checked. If there is a pair of consecutive keywords that occurs in the text, they will be combined to a term. After that, they will be added to corresponding terms list. In worse cases that there is no pair of keywords found, those keywords which are stop-word will be removed. The others will be included in terms list. At the second step, the algorithm tries to extract all the known words (e.g. product name, product type, etc.). These words will be added to Known Terms list.

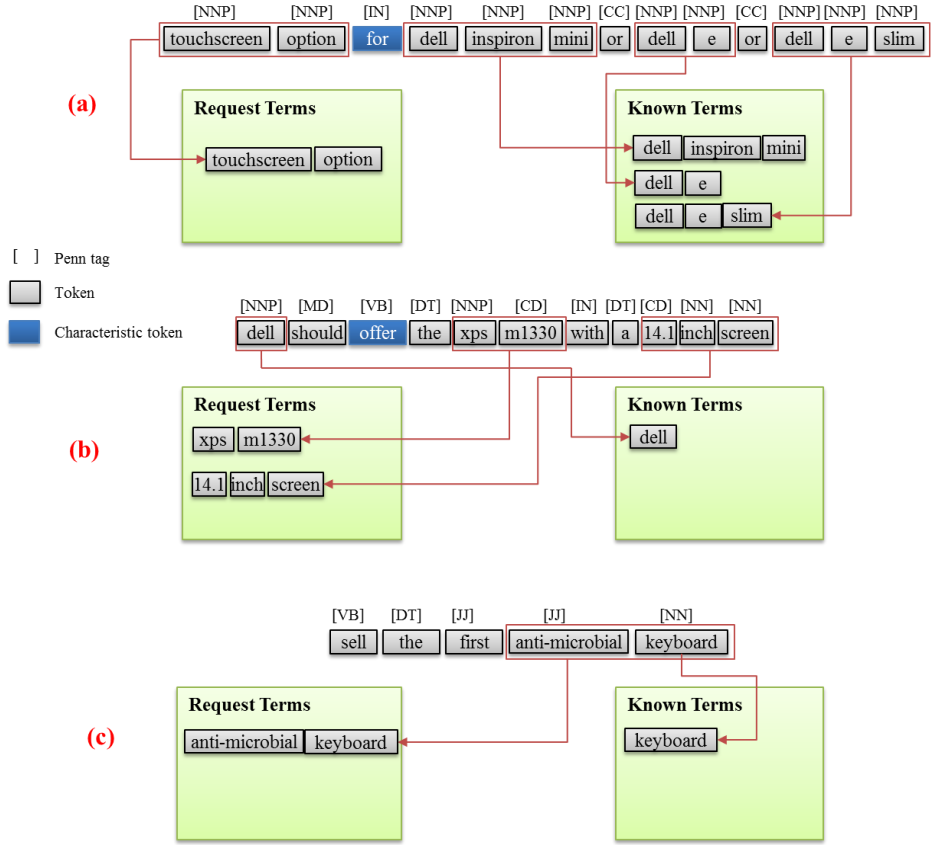

Fig 3. Illustration of Algorithm 1. (a) Case 1 is active; (b) Case 2 is active; (c) General case

---

**Algorithm 2** ExtractWholeIdea($I$)

    **Input**: an idea text $I$, $I = \{s_1, \ldots, s_n\}$ /* Definition 1 */
    **Output**: terms list $V$

1.    $KT \leftarrow \{\varnothing\}$
2.    $RT \leftarrow \{\varnothing\}$
3.    $H^k \leftarrow \{\varnothing\}$    /* Create an empty hash table of keywords */
4.    $H^n \leftarrow \{\varnothing\}$    /* Create an empty hash table common words */
5.    $tmp \leftarrow \{\varnothing\}$
6.    $str \leftarrow \text{removeHTMLtags}(I)$    /* Remove all HTML tags */
7.    $t \leftarrow \text{tokenize}(str)$    /* Tokenize $str$ into single lower case words and store into array $t$ */
8.    $t \leftarrow \text{removeStopword}(t)$    /* Remove all stop words */
9.    $t \leftarrow \text{filterByLength}(t)$    /* Remove words that have number of characters smaller than 2 */
10.   $t \leftarrow \text{stemming}(t)$    /* Apply stemming using Porter algorithm [59] */
11.   **FOR** $i \leftarrow 1$ **TO** $|t|$ **DO**
12.      $score \leftarrow wf(t_i)$    /* compute word frequency $wf$ of $t_i$ */
13.      **IF** $score > 1$ **THEN** $H^k \leftarrow (t_i, i)$    /* $t_i$ is a *key* and $i$ is *value* associated with $t_i$ */
14.      **ELSE** $H^n \leftarrow (t_i, i)$
15.   **ENDFOR**
16.   **FOR** each pair $(h_i, h_j)$ in $H^k$ **DO**
17.      **IF** $\text{abs}(\text{value}(h_i) - \text{value}(h_j)) = 1$ **THEN** /* abs($X$) returns absolute value of X */
18.        **IF** $\text{value}(h_i) < \text{value}(h_j)$ **THEN** $tmp \leftarrow \text{concatnate}(\text{key}(h_i), \text{key}(h_j))$
19.        **ELSE** $tmp \leftarrow \text{concatnate}(\text{key}(h_j), \text{key}(h_i))$

```
20      IF tmp ∈ K THEN KT ← tmp ELSE RT ← tmp
21      tmp ← {∅}
22    ENDIF
23   ENDFOR
24   FOR each h ∈ H^n DO
25     IF h ∈ K THEN KT ← h
26   ENDFOR
27   RETURN V ← RT, KT
```

Fig. 4 illustrates an example (Ex2) of extracting terms using Algorithm 2. After removing stop-words "on", the pairs of keywords "color" – "desktop" and "notebook" – "desktop" in the text become consecutive words. Notice that "color" – "notebook" is not consecutive, hence there is no term created from this pair.

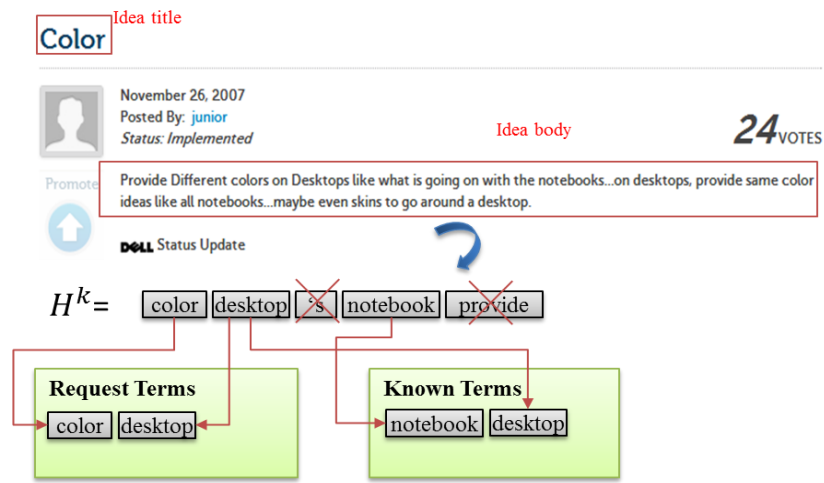

Fig 4. Illustration of Algorithm 2 (Ex2)

Before we get the final terms list, the *RT* and *KT* list are refined. In Fig 3(b), the term "xps 1330" is a known term; hence it is moved to *KT* list. When Algorithm 1 is performed, it often generates an empty *KT* list. To avoid this problem, a procedure of extracting all known words in Algorithm 2 is used. In worst case, there is still possibility of *RT* and *KT*'s being empty.

**3.1.3 Relevance measurement**

In order to reflect the *interest level* of an *RT*, we define a function *tr(RT)* representing its weight value for measuring how trendy *RT* is. To retrieve *tr(RT)*, a scoring method based on Google Insight (GI) [12] for Search is proposed. GI is a web service of Google, which analyzes a portion of Google web searches to compute how many searches have been done for a given term, relative to total number of searches done on Google over time. Our scoring method is as follows:

Suppose, given a term *T*, *d* times of search result for *T* has been returned from GI. Then, the trend of *T* is described as the average number of searches done on the Google service in *d* units of time. Thus, if $\alpha_i$ expresses an insight value returned by GI at a data point, then:

$$tr(T) = \frac{\sum_{i=1}^{d} \alpha_i}{d} \quad (3.1)$$

From Ex1, we want to get an interest level of the term "anti-microbial keyboard" in time range between May, 2006 and May, 2007 (the idea associated with the term was posted on June 5[th], 2007). Table 2 shows the results obtained by GI. Sometimes, we need to refine the term to retrieve GI based interest level. Notice that we used refined query "anti-microbial" instead of using "anti-microbial keyboard".

Table 2. Result from GI for Search the query ("anti-microbial")

| Time | $\alpha_i$ |
| --- | --- |
| 2006-09 | 57 |
| 2006-10 | 75 |
| 2006-11 | 94 |
| 2006-12 | 100 |
| 2007-01 | 88 |
| 2007-02 | 64 |
| 2007-03 | 72 |
| 2007-04 | 67 |
| 2007-05 | 72 |

By using Equation 3.1, and the result in Table 2, if we set d=1, *tr*("anti-microbial keyboard") = 72 as of the last data point, 2007-05.

Now, we have to consider the *scope of change*. According to Huang *et al.* [6], an idea is potentially implemented when the *scope of change* of an idea is relatively low. For a known term, KT, we define a function *sp*(KT) to represent *scope of change*, and it is usually pre-defined. Scope of the change for $KT = \{t_i | i=1,\ldots,J\}$ is defined as

$$sp(KT) = \begin{cases} w(t_j) & J = 1 \\ \min\{w(t_j) | j=1,\ldots,J\} & J > 1 \end{cases} \quad (3.2)$$

From Ex1, the weight value associated with "keyboard" is 3, and we have *sp*("keyboard") = 3. Similarly, for the known term "Dell XCD35", we have *w*("Dell") = 4, *w*("XCD35") = 1, and *sp*("Dell XCD35") = 1.

To check the balance of a terms list, we introduce the *well-balanced measurement*, which is adopted from [51]. Given a terms list *V*(α, β), let $p = |\alpha \cup \beta|$ be the number of all terms (request and known) in *V*, and $q = |\alpha|$ be the number of request terms. Then, we define *b(V)* as measure for well-balanced distribution of *V* as bellows.

$$b(V) = \begin{cases} \dfrac{2 \times (p-q)}{p} & (q \geq \dfrac{p}{2}) \\ \dfrac{2 \times q}{p} & (q < \dfrac{p}{2}) \end{cases} \quad (3.3)$$

From Ex1, *V*(α, β) = ({"anti-microbial keyboard"}, {"keyboard"}), and *b(V)* = 1

Finally the definition of relevance score is provided in Definition 9.

**Definition 9 (Relevance score)**
The relevance score of an idea is to measure how this idea is relevant to the firm. Let *rel*(*I*) be the relevance score of an idea with terms list *V*(α, β). Then:

$$rel(I) = \frac{|\beta|}{|\alpha|} \times \frac{\sum_{k=1}^{|\alpha|} tr(\alpha_k)}{\sum_{k=1}^{|\beta|} sp(\beta_k)} \times \ln(c + b(V)) \quad (3.4)$$

The factor $\ln(c + b(V))$ is used to adjust *b(V)* in case of small *b(V)*. In this paper, we choose *c* = 1.72 because ln(2.72) is approximated to 1.

From Ex1, for *V*(α, β) = ({"anti-microbial keyboard"}, {"keyboard"}), we have: *tr*("anti-microbial keyboard") = 30, *sp*("keyboard") = 3 and *b(V)* = 1. Therefore, *rel*(*I*) = 10.00632.

**3.2 Interest evaluation**

**3.2.1 Retrieve crowd information**

This section explains how to calculate interest score, that is evaluated from the number of votes on the idea, the profile of users who gave comments on the idea, the profile of the owner of the idea, and information associated with the comments (e.g. comment's date, number of comments). Fig. 5 shows an example idea (Ex3) from Dell IdeaStorm website, and comments about the idea which we can retrieve from the crowdsourcing application.

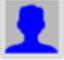

Fig. 5. An example (Ex3) idea from Dell IdeaStorm website

### 3.2.2 Interest measurements

In this section, five measurements for interest evaluation are defined as follows.

**Definition 10 (Vote)**
A function *vote*(*I*) is defined to return the number of votes on an idea *I*, which means the number of users interested on this idea. It can be obtained directly from the idea's page shown in Fig. 5.

**Definition 11 (User type)**
A function *type*(*I*) returns the type of the user who posts the idea *I*. According to [34, 35], the ideas proposed by users who have past experience of success, significantly have higher chance to be implemented than others. Such users are called "serial users". We adopt this insight into our model. Hence *type*(*I*) returns binary value of 1 for a serial user, 0 for a non-serial user.

**Definition 12 (Diversity degree)**
According to diversity criteria from [5], the function *div*(*I*) returns the diversity degree of users who comment on the idea *I*. Let *n* be the number of distinct users who comment on the idea, and $p_i$ is the proportion of comments from *i*-th user. The diversity degree is determined by entropy function:

$$\mathrm{div}(I) = \begin{cases} -\sum_{i=1}^{n} p_i \log_2 p_i & n \geq 1 \\ -1 & n < 1 \end{cases} \qquad (3.5)$$

**Definition 13 (Concern degree)**
The function *con*(*I*) returns the average interval between comments on the idea *I*. It presents the concern of the crowd over time on the idea *I*. As shown in Fig. 5, it can be obtained from the comment page of idea *I*. The unit of this measurement is number of days. Suppose idea *I* receives the first comment on date *fd*, its last comment is on date *ed* and there is *n* comments. Concern degree is described as:

$$\mathrm{con}(I) = \begin{cases} -1 & n \leq 1 \\ \dfrac{ed - fd}{n - 1} & n > 1 \end{cases} \qquad (3.6)$$

**Definition 14 (Expert's interest)**
The function *epr*(*I*) is used to return the number of comments from the experts in the company. This measurement presents how NPD experts follow up on the idea *I*. As in Fig. 5, it can be counted from the comment page of idea *I*.

From Ex 3, five measurements of interest of idea are calculated as below:

$\mathrm{vote}(I) = 26$

$\mathrm{type}(I) = 0$

$\mathrm{div}(I) = -(0.5\log_2 0.5 + 0.5\log_2 0.5) = 1$

$\mathrm{con}(I) = 0$

epr(*I*) = 1

## 3.3 Measurement of probability of being potential

Using the relevance measurements and the interest measurements, we finally estimate an idea's probability of being potential. Utilized for this purpose is a binary classifier with which the value 1 represents "Potential," and the value 0, "Not potential". Hence, for an idea *I*, our purpose is to calculate the probability *p*(*I*) of idea *I* given a set of input variables :

$$p = \Pr\big(\text{potential}(I) = 1 | \text{rel}(I), \text{vote}(I), \text{type}(I), \text{epr}(I), \text{div}(I), \text{con}(I)\big) \quad (3.7)$$

The probability is acquired based on Logistic regression [47], which is a well-known technique to estimate the probability of an output variable from a set of input variables. Table 3 shows the main notations used in this section:

Table 3. Notation in section 3.3

| Notation | |
|---|---|
| $x$ | vector of input variables (attributes) |
| $y$ | output variable. (1: if an idea is potential, 0: otherwise) |
| $p$ | probability of being potential, $0 \leq p \leq 1$ |
| $\beta$ | vector of weight value associated with $x$, $|\beta| = |x|$ |
| $\hat{\beta}$ | vector of maximum likelihood (ML) estimate of $\beta$ |
| $D$ | set of labeled data (ideas) consists a 2-tuple $(x, y)$ |
| $B$ | set of parameter vectors, $B = \{\beta_1, \beta_2, \ldots, \beta_{|B|}\}$ |
| $W$ | set of associated weight of $B$, $W = \{w_1, w_2, \ldots, w_{|B|}\}$ |
| $\omega$ | a reduction function $\omega = e^{-\alpha}, \alpha > 0$ |
| $\varepsilon$ | Threshold value $\varepsilon \in \square^+$ |
| $\sigma$ | Minimum relative error improvement $\sigma \in \square^+$ |
| $\eta_0$ | Initial learning rate $\eta_0 \in \square^+$ |
| $\delta$ | Annealing rate $\delta \in \square^+$ |
| $\Omega$ | Set of observed instances |
| $\theta$ | Maximum size of *B*, $\theta \in \square^+$ |

For an arbitrary idea *I*, considering its prediction *p*, using logistic regression we have:

$$p = \frac{e^{(\beta^T x)}}{1 + e^{(\beta^T x)}} \quad (3.8)$$

Where $\beta = \langle \beta_0, \beta_1, \ldots, \beta_6 \rangle$ and $x = \langle 1, \text{rel}(I), \text{vote}(I), \text{type}(I), \text{epr}(I), \text{div}(I), \text{con}(I) \rangle$.

Suppose we are given a labeled data set $D = \{(x_1, y_1), \ldots, (x_i, y_i), \ldots, (x_n, y_n)\}$, from this our purpose is to learn the logistic regression parameter vector $\beta$ for Equation (3.8). It means we need to estimate:

$$\hat{\beta} = \arg\max_{\beta} L(D | \beta) = \arg\max_{\beta} \sum_{i=1}^{n} \log \Pr(y = y_i | x_i, \beta) \quad (3.9)$$

In order to solve Equation (3.9), we can consider employing traditional logistic regression methods or other learning methods such as neural network, support vector machine, etc. However, traditional methods are batch learning algorithms, that is, the vector $\hat{\beta}$ is fixed overtime. In this paper, we develop an online learning algorithm instead of using traditional ones. It means the vector $\hat{\beta}$ will be updated after a new instance of labeled data arrived. Our purpose is to find an update rule *f* that allows us to bound the sum of losses as in Equation (3.9). For this purpose, we employed the stochastic gradient descent (SGD) rule which was introduced in [61, 62]. However, because SGD was designed for batch learning, we extended its applicability to online mode learning by combining it with the weighted majority mechanism [63] as shown in Algorithm 3. For each new instance, the algorithm samples a parameter vector $\beta_r$ randomly from *B*. For a pre-defined threshold value, if the prediction on the considered instance using $\beta_r$ incurs a loss value that is greater than or equal to the threshold value, the algorithm learns a new $\hat{\beta}$ according to SGD. After that, if the size of $B$ does not exceed the maximum size, the algorithm adds $\hat{\beta}$ to $B$. A new weight associated with $\hat{\beta}$ also is added to $W$. In addition, the algorithm decreases the weight value of $\beta_r$ by an amount of $\omega$.

---

**Algorithm 3** OLR $(x, y, \varepsilon, \sigma, \eta_0, \delta, \Omega, B, W, \omega, \theta)$

    **Input**: $x, y, \varepsilon, \sigma, \eta_0, \delta, \Omega, B, W, \omega, \theta$
    **Output**: $p, B, W, \Omega$
1    $\Omega \leftarrow (x, y)$ /* add $x, y$ to the set of observed instances */
2    $H \leftarrow \sum_{i=1}^{|B|} w_i$ /* $w_i$ returns the weight associated with parameter $\beta_i$ */
3    $\beta_r \leftarrow \text{sampling}\left(\frac{w_1}{H}, \frac{w_2}{H}, \ldots, \frac{w_{|B|}}{H}\right)$ /* sample a random $\beta_r$, $r$ is its index in $B$ */
4    $\ell \leftarrow 0$ /* initiate loss value */
5    $\hat{\ell} \leftarrow \infty$
6    $t \leftarrow 0$ /* initiate trial value */
7    $p \leftarrow \dfrac{e^{(\beta_r^T x)}}{1 + e^{(\beta_r^T x)}}$ /* Calculate probability of being potential */
8    **IF** $\text{abs}\left(\log_2\left(p^y (1-p)^{(1-y)}\right)\right) \geq \varepsilon$ **THEN** /* The prediction makes mistake */
9      $\eta_e \leftarrow \eta_0 / (1 + e/\delta)$
10    $\hat{\beta} \leftarrow \beta_r$ /* initiate a new parameter */
11    $w_r \leftarrow w_r \times \omega$ /* update weight value of $\beta_r$ */
12    **IF** $|B| < \theta$ **THEN**
13      **WHILE** $\text{relDiff}(\hat{\ell}, \ell) > \sigma$ /* Define $\text{relDiff}(a, b) = \dfrac{\text{abs}(a-b)}{\text{abs}(a) + \text{abs}(b)}$ */
14        **FOR** $i \leftarrow 1$ **TO** $|\Omega|$ **DO**
15          $p_i = \left(e^{(\hat{\beta} \cdot x_i)}\right) / \left(1 + e^{(\hat{\beta} \cdot x_i)}\right)$
16          $\hat{\beta} \leftarrow \hat{\beta} + \eta_e x_i \left(I(y_i = 1) - p_i\right)$ /* $I(a = 1)$ returns 1 if $a = 1$, otherwise 0 */
17        **ENDFOR**
18        $\hat{\ell} \leftarrow \ell$
19        $\ell \leftarrow -\sum_{i \leq |\Omega|} \log_2 \left(p_i^{y_i} (1-p_i)^{(1-y_i)}\right)$
20        $t \leftarrow t + 1$
21      **ENDWHILE**
22      $B \leftarrow \hat{\beta}$
23      $W \leftarrow 1$
24    **ENDIF**
25  **ENDIF**
26  **RETURN** $p, B, W, \Omega$

---

In order to analyze the performance of a learning algorithm, its regret bound is estimated. The learning algorithm's regret is the difference between its number of mistakes and the number of mistake the optimal parameter vector $\beta^*$ in $B$ makes on the same sequence of labeled data. We present the regret bound of our OLR algorithm.

**Theorem 1** Suppose we have a set of parameter vectors $B = \{\beta_1, \beta_2, \ldots, \beta_s\}$. Let $D = \{(x_1, y_1), (x_2, y_2), \ldots, (x_{i^*}, y_{i^*}), \ldots, (x_n, y_n)\}$ be an arbitrary set of labeled data. Also suppose the prediction by OLR on $(i^* - 1)$-th instance makes a mistake which incurs an increase in the size of $B$ such that $(s+1) = \theta$. If the reduction function is $\omega = e^{-\alpha}$, the theory regret $\Re$ of OLR on $D$ is described as:

$$\Re \leq \theta - 1 + \frac{\alpha(n - \theta + 1)}{8} + \frac{\ln\left(e^{-\alpha(\theta-1)} + \theta - 1\right)}{\alpha} \quad (3.10)$$

The proof of Theorem 1 is given in Appendix.

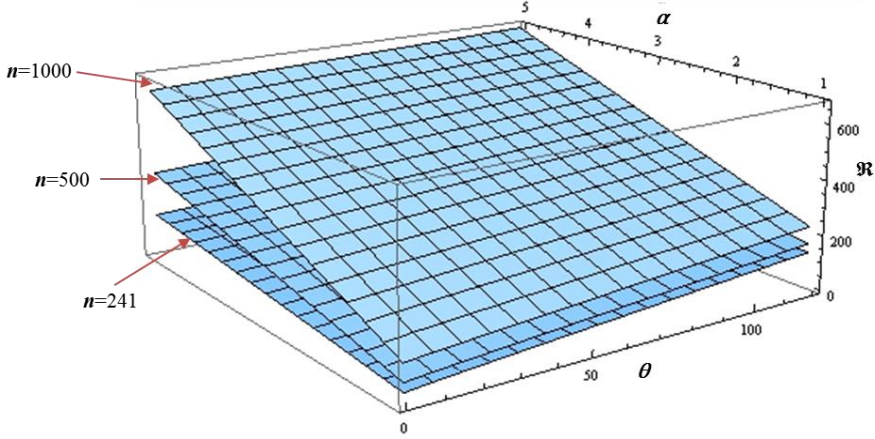

Fig. 6. Illustration of theory regret

The right-hand side of Equation (3.10) is a function with three input variables: the number of instances $n$, the parameter $\alpha$ of reduction function, and $\theta$ - the maximum size of set of parameter vectors. Fig. 6 illustrates the value of theory regret with varying input variables. As $n$ and $\alpha$ are increased, the theory regret $\Re$ rapidly increases. Contrarily, $\theta$ does not have high impact on the increase of $\Re$. Therefore, $\alpha$ should be assigned small value. The value of $\theta$ can be varied from 30 to 120.

## 4. Numerical experiments

In this section, we illustrate the performance of our methods by conducting experiments of comparing the OLR algorithm with two well-known learning algorithms based on the scheme of RapidMiner [64], which are support vector machine (SVM) and traditional logistic regression (TLR). Note that, both two algorithms are batch learning algorithms.

### 4.1 Experiment's setting

Data for this paper come from the public textual information on Dell's IdeaStorm website, which are opened to all people. Fig. 7 shows an overall procedure of selecting idea from crowd. After registration, individuals are allowed to post relevant ideas; they can either promote or demote others' ideas (voting points) by commenting on any ideas and express their opinions in more detail in an online voting system. Some expert users in the company maintain the website by passing the relevant ideas to corresponding internal groups for review, and they can communicate with the idea posting individual to comment about the idea. Based on the communication result, they change the status of the idea.

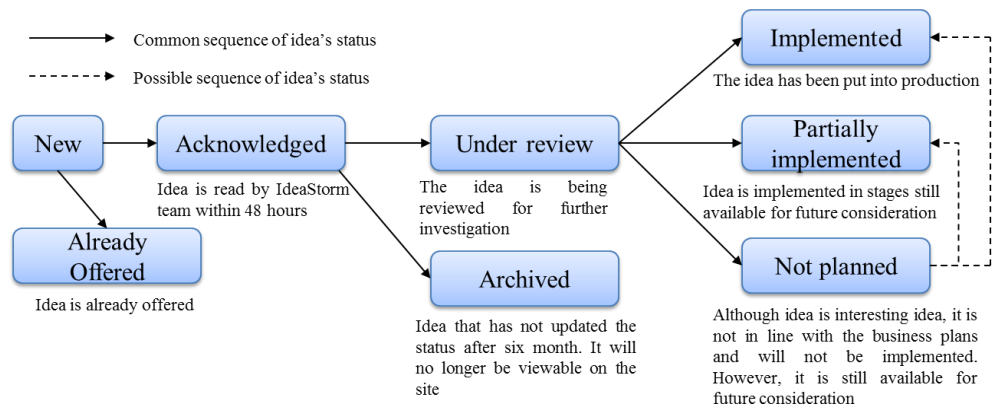

Fig. 7. The evolution of idea's status in IdeaStorm.

Every idea starts with "New" state upon submission. Most of the posted ideas received "Acknowledge" within 48 hours. If the expert users find that an idea is already part of a product or service, they will update the status to "Already Offered". Among remaining ideas, the expert users pass some to appropriate business team for review, and they change the status to "Under review". After carefully evaluating these ideas, Dell makes one of three decisions: "Implemented", "Partially implemented" or "Not Planned". Once an idea is "Implemented", it is closed for votes and comments. "Partially implemented" and "Not Planned" ideas are still available for future consideration, thus the individuals can continue vote and comment on these ideas. Ideas that do not receive any concern of either individuals or expert users after six month are "Archived" and thus no longer be viewable on the site.

Table 4 shows three categories of all ideas. When an individual posts an idea on IdeaStorm, this person selects the Category as well as the sub categories to which the idea belongs.

Table 4. Idea's categories gathered from Dell's IdeaStorm

| Categories | Sub categories |
|---|---|
| Dell Ideas | Advertising and Marketing; Dell Community; Dell Web Site; IdeaStorm; Retail; Service and Support. |
| Topic Ideas | Education; Enterprise; Environment; Gaming; Healthcare and Life Sciences; Small Business; Storm Session Topics; Women's Interest |
| Product Ideas | Accessories (Keyboards, etc.); Alienware; Broadband and Mobility; Desktops and Laptops; Mobile Devices; New Product Ideas; Operating Systems; Printers and Ink; Servers and Storage; Software |

Because this paper focuses on NPD, we only use ideas from the last category (Product Ideas), and sub categories Printers and Ink, Servers and Storage and Software are excluded since these two sub categories do not contribute much to NPD of Dell. We collected and analyzed 7,090 ideas that were posted from 14 February 2007 to 13 November 2011. In this paper, ideas that are in any of "Implemented", "Partially implemented", and "Under Review" are considered "Potential". "Archived" and "Not Planned" ideas are considered "Not potential". We do not take "New" ideas and "Acknowledged" ideas into account since the final decision on these ideas is not known. For the rest of "Potential" and "Not Potential" ideas, we removed all spam ideas and non-relevant ideas. Finally, there are 77 "Implemented" ideas, 48 "Partially Implemented" ideas, 7 "Under Review" ideas, 29 "Not Planned" ideas and 81 "Archived" ideas remained. Thus, we have 132 "Potential" ideas and 110 "Not Potential" ideas, which are used for our experiments.

For the scope measurement, all terms that have the same scope level received the same weight value. In order to illustrate the effect of weight values for scope measurement, we used three settings. For Setting 1, we defined weight value of each scope level as in Table 5.

Table 5. Weight value for scope measurement in setting 1

| Scope | Description | Weight |
|---|---|---|
| Product | Terms that related to specific products, e.g. Alienware M11x, Latitude E4200 | 1 |
| Product line | Terms that related family of product, e.g. Alienware, Latitude, Inspirion | 2 |
| General case | Terms that are general expression, e.g. laptop, notebook, printer, monitor, etc. | 3 |
| Unknown | Terms that not related to Dell included the word "dell" | 4 |

For the scope level of setting 2 and 3, we assigned a random number following discrete uniform distribution as shown in Table 6. For each test, we ran 30 trials and collected its statistics.

Table 6. Weight value for scope measurement in setting 2 and setting 3

| Scope | Weight | |
|---|---|---|
| | Setting 2 | Setting 3 |
| Product | Discrete Uniform(1,10) | Discrete Uniform(1,25) |
| Product line | Discrete Uniform(11,20) | Discrete Uniform(26,50) |
| General case | Discrete Uniform(21,30) | Discrete Uniform(51,75) |
| Unknown | Discrete Uniform(31,40) | Discrete Uniform(76,100) |

For the trend measurement, we used GIS and set $d = 1$, which means we obtained the closest data point (month) to the time that the idea was published. It is important to note that, Google does not provide free API to access GIS data automatically, thus we have to perform searches for the interest degree on GIS manually. We conducted the searches on June 11$^{th}$, 2012. The other measurements in Section 4.2 can be simply obtained by using data mining techniques.

**4.2 Validation of the proposed text mining technique**

After evaluating the measurements mentioned in Section 3.1 and 3.2, we have a data table which have 10 columns and 242 rows. Table 7 illustrates an example of 10 rows from the data table. The last column "label" is the true decision from Dell's expert user, 0 means "Not Potential" and 1 means "Potential".

Table 7. Example of data table after evaluating measurements

| Idea ID | Refined RT | Refined KT | rel | vote | type | div | con | epr | True label |
|---|---|---|---|---|---|---|---|---|---|
| 1 | 15.4 inch | xps; | 27.02 | 262 | 0 | 4.82 | 7.67 | 1 | 1 |
| 2 | plastic shell; metal casing; | notebook; | 20.12 | 1039 | 0 | 5.27 | 35.91 | 0 | 1 |
| 3 | web cam; microphone; | laptop | 22.41 | 1833 | 0 | 7.81 | 4.38 | 1 | 1 |
| 4 | dvd jukebox | jukebox | 55.64 | -11 | 0 | 1.58 | 253.00 | 0 | 0 |
| 5 | tablet pc | e1405; m1210; xps; notebook; | 34.22 | 1138 | 0 | 8.68 | 2.17 | 2 | 1 |
| 6 | Internet Linux desktop | desktop | 33.35 | 4729 | 0 | 7.44 | 4.61 | 3 | 0 |
| 7 | esata port; | notebook; | 33.35 | 181 | 0 | 5.61 | 14.88 | 4 | 1 |
| 8 | light laptop; | laptop | 22.75 | 323 | 0 | 4.66 | 23.70 | 2 | 1 |
| 9 | rugged Laptop | laptop | 21.21 | 16 | 0 | 1.58 | 373.50 | 2 | 1 |
| 10 | silent computer; quite computer; | computer | 14.21 | 3666 | 0 | 8.32 | 11.24 | 1 | 0 |

The result of the proposed text mining method has been evaluated by conducting a survey. Google Form service was used to publish our survey. In proportion to the number of ideas, the survey consists of 242 questions. For each question, the participants were asked to give their opinion whether they "Totally agree", "Agree", "Neutral", "Partially not agree" or "Not agree" with the extraction of Request Terms and Known Terms. For most of the ideas, over 50% of observations show that they agreed with the results of the quality of RTs and KTs extracted, and only 19.7% of observations did not agree with the results. Fig. 8 shows the total results of the surveys.

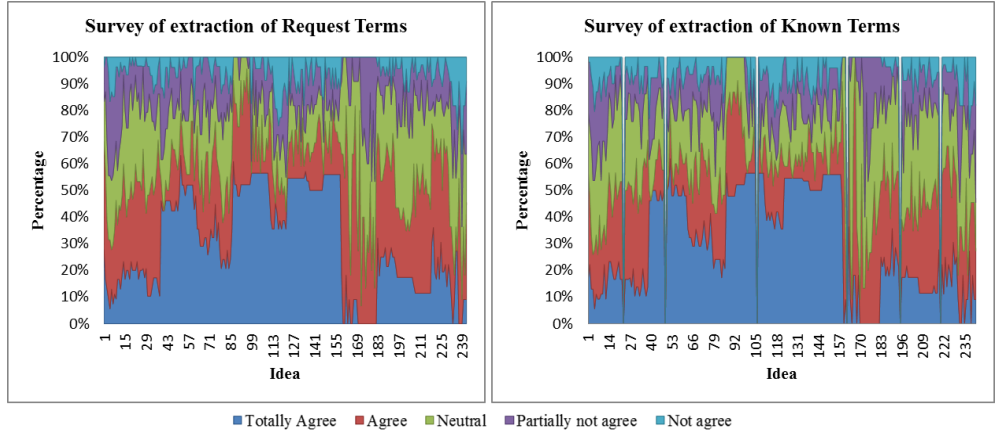

Fig. 8. Results from survey of the proposed text mining method

**4.3 Performance of OLR**

Before applying the learning algorithm, we test the significance of the 6 input variables (rel, vote, type, div, con and epr). The test was set at 90% confidence interval for all three settings of scope measurement. Table 8 shows the result of *p*-value of this test.

Table 8. Logistic regression's significance test results

| Input variable | *p*-value | | | | |
| --- | --- | --- | --- | --- | --- |
| | Setting 1 | Setting 2 | | Setting 3 | |
| | | Mean | Var. | Mean | Var. |
| Intercept | 0.000 | 0.000 | 0.000 | 0.000 | 0.000 |
| rel | 0.046 | 0.278 | 0.102 | 0.415 | 0.099 |
| type | 0.062 | 0.066 | 0.000 | 0.071 | 0.000 |
| con | 0.238 | 0.222 | 0.001 | 0.212 | 0.000 |
| epr | 0.000 | 0.000 | 0.000 | 0.000 | 0.000 |
| vote | 0.106 | 0.108 | 0.000 | 0.100 | 0.000 |
| div | 0.072 | 0.074 | 0.000 | 0.071 | 0.000 |

As shown in Table 8, the relevance, type, diversity and expert's interest measurements are significant. Even though vote and concern degree are not so significant, they are still acceptable to use in our learning model. In addition, the results of Setting 2 and Setting 3 show that if we assign large weight value for each scope level, it would reduce the significance of relevance measurement. Thus, we use Setting 1 for the learning algorithm.

For the learning algorithm, we varied the threshold value $\varepsilon$, parameter *α* in the reduction function *ω*, and the maximum size $\theta$ of the set of parameters *B* in order to examine how the result changes depending on these values. As shown in Table 9, we have 75 sets of combination to examine. For each set we ran 30 trials. Each trial consists of 241 instances of idea (one instance is sampled randomly to train the first parameter vector). These instances are arranged in a random sequence for each trial.

Table 9. Combinations of $\varepsilon$, *α,* and $\theta$

| Set No. | $\varepsilon$ | α | $\theta$ | Set No. | $\varepsilon$ | α | $\theta$ | Set No. | $\varepsilon$ | α | $\theta$ |
| --- | --- | --- | --- | --- | --- | --- | --- | --- | --- | --- | --- |
| 1 | 0.1 | 1 | 30 | 26 | 0.3 | 3 | 30 | 51 | 0.7 | 2 | 30 |
| 2 | 0.1 | 1 | 50 | 27 | 0.3 | 3 | 50 | 52 | 0.7 | 2 | 50 |
| 3 | 0.1 | 1 | 80 | 28 | 0.3 | 3 | 80 | 53 | 0.7 | 2 | 80 |
| 4 | 0.1 | 1 | 100 | 29 | 0.3 | 3 | 100 | 54 | 0.7 | 2 | 100 |
| 5 | 0.1 | 1 | 120 | 30 | 0.3 | 3 | 120 | 55 | 0.7 | 2 | 120 |
| 6 | 0.1 | 2 | 30 | 31 | 0.5 | 1 | 30 | 56 | 0.7 | 3 | 30 |
| 7 | 0.1 | 2 | 50 | 32 | 0.5 | 1 | 50 | 57 | 0.7 | 3 | 50 |
| 8 | 0.1 | 2 | 80 | 33 | 0.5 | 1 | 80 | 58 | 0.7 | 3 | 80 |
| 9 | 0.1 | 2 | 100 | 34 | 0.5 | 1 | 100 | 59 | 0.7 | 3 | 100 |
| 10 | 0.1 | 2 | 120 | 35 | 0.5 | 1 | 120 | 60 | 0.7 | 3 | 120 |
| 11 | 0.1 | 3 | 30 | 36 | 0.5 | 2 | 30 | 61 | 1 | 1 | 30 |
| 12 | 0.1 | 3 | 50 | 37 | 0.5 | 2 | 50 | 62 | 1 | 1 | 50 |
| 13 | 0.1 | 3 | 80 | 38 | 0.5 | 2 | 80 | 63 | 1 | 1 | 80 |
| 14 | 0.1 | 3 | 100 | 39 | 0.5 | 2 | 100 | 64 | 1 | 1 | 100 |

| 15 | 0.1 | 3 | 120 | 40 | 0.5 | 2 | 120 | 65 | 1 | 1 | 120 |
| 16 | 0.3 | 1 | 30  | 41 | 0.5 | 3 | 30  | 66 | 1 | 2 | 30  |
| 17 | 0.3 | 1 | 50  | 42 | 0.5 | 3 | 50  | 67 | 1 | 2 | 50  |
| 18 | 0.3 | 1 | 80  | 43 | 0.5 | 3 | 80  | 68 | 1 | 2 | 80  |
| 19 | 0.3 | 1 | 100 | 44 | 0.5 | 3 | 100 | 69 | 1 | 2 | 100 |
| 20 | 0.3 | 1 | 120 | 45 | 0.5 | 3 | 120 | 70 | 1 | 2 | 120 |
| 21 | 0.3 | 2 | 30  | 46 | 0.7 | 1 | 30  | 71 | 1 | 3 | 30  |
| 22 | 0.3 | 2 | 50  | 47 | 0.7 | 1 | 50  | 72 | 1 | 3 | 50  |
| 23 | 0.3 | 2 | 80  | 48 | 0.7 | 1 | 80  | 73 | 1 | 3 | 80  |
| 24 | 0.3 | 2 | 100 | 49 | 0.7 | 1 | 100 | 74 | 1 | 3 | 100 |
| 25 | 0.3 | 2 | 120 | 50 | 0.7 | 1 | 120 | 75 | 1 | 3 | 120 |

In table 10, we obtained mean, variance of Accuracy, Precision and Recall of the result. Because the proposed prediction model is an online learning algorithm, it is also necessarily to examine the elapsed time of learning, which is also shown at the last column of Table 10.

Table 10. Best results of the proposed prediction model

| Set No. | $\varepsilon$ | $\alpha$ | $\theta$ | Accuracy | | Precision | | Recall | | Elapsed (ms) |
|---|---|---|---|---|---|---|---|---|---|---|
| | | | | Mean | Std. | Mean | Std. | Mean | Std. | |
| 34 | 0.5 | 1 | 100 | 0.66 | 0.04 | 0.68 | 0.03 | 0.70 | 0.06 | 97.58 |
| 38 | 0.5 | 2 | 80  | 0.66 | 0.03 | 0.69 | 0.03 | 0.68 | 0.06 | 90.56 |
| 40 | 0.5 | 2 | 120 | 0.66 | 0.04 | 0.69 | 0.04 | 0.68 | 0.05 | 107.23 |
| 48 | 0.7 | 1 | 80  | 0.66 | 0.03 | 0.69 | 0.03 | 0.69 | 0.04 | 95.88 |
| 49 | 0.7 | 1 | 100 | 0.66 | 0.02 | 0.69 | 0.03 | 0.69 | 0.06 | 90.61 |
| **53** | **0.7** | **2** | **80** | **0.66** | **0.04** | **0.69** | **0.04** | **0.68** | **0.06** | **83.34** |
| 55 | 0.7 | 2 | 120 | 0.66 | 0.04 | 0.69 | 0.03 | 0.68 | 0.06 | 104.67 |
| **58** | **0.7** | **3** | **80** | **0.66** | **0.04** | **0.69** | **0.03** | **0.69** | **0.07** | **80.59** |
| 59 | 0.7 | 3 | 100 | 0.66 | 0.03 | 0.69 | 0.03 | 0.68 | 0.06 | 95.63 |
| 63 | 1 | 1 | 80  | 0.66 | 0.03 | 0.69 | 0.03 | 0.68 | 0.06 | 99.61 |
| 64 | 1 | 1 | 100 | 0.67 | 0.03 | 0.70 | 0.03 | 0.70 | 0.06 | 111.93 |
| 65 | 1 | 1 | 120 | 0.66 | 0.04 | 0.69 | 0.04 | 0.68 | 0.06 | 103.72 |
| **67** | **1** | **2** | **50** | **0.66** | **0.04** | **0.69** | **0.04** | **0.68** | **0.05** | **43.85** |
| 68 | 1 | 2 | 80  | 0.66 | 0.03 | 0.69 | 0.02 | 0.69 | 0.05 | 90.63 |
| 69 | 1 | 2 | 100 | 0.66 | 0.04 | 0.69 | 0.03 | 0.68 | 0.06 | 101.12 |
| 73 | 1 | 3 | 80  | 0.66 | 0.03 | 0.70 | 0.03 | 0.68 | 0.05 | 92.65 |
| **74** | **1** | **3** | **100** | **0.67** | **0.03** | **0.70** | **0.03** | **0.69** | **0.06** | **88.15** |
| 75 | 1 | 3 | 120 | 0.66 | 0.03 | 0.70 | 0.03 | 0.68 | 0.05 | 95.24 |

Statistics of the resulting accuracy, precision, and recall are plotted in Fig. 9, Fig. 10, and Fig. 11 respectively.

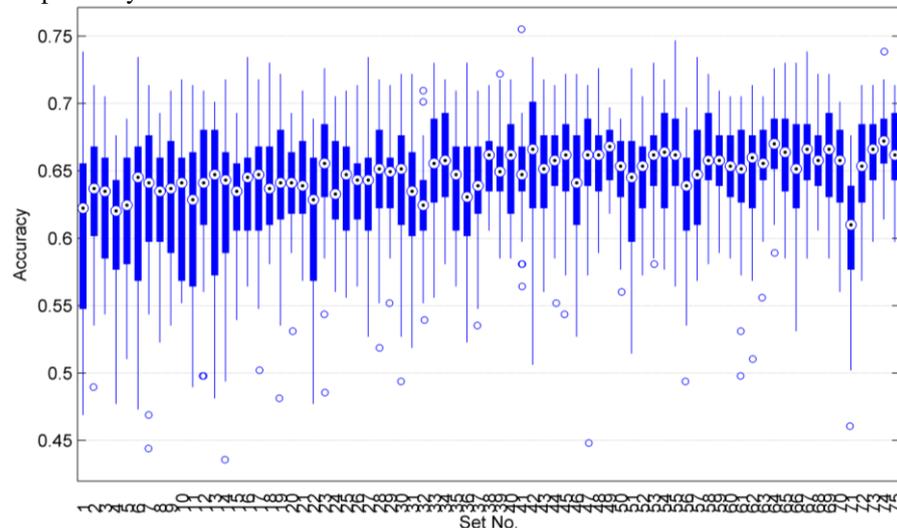

Fig. 9. Box plot of the accuracy of the proposed prediction model

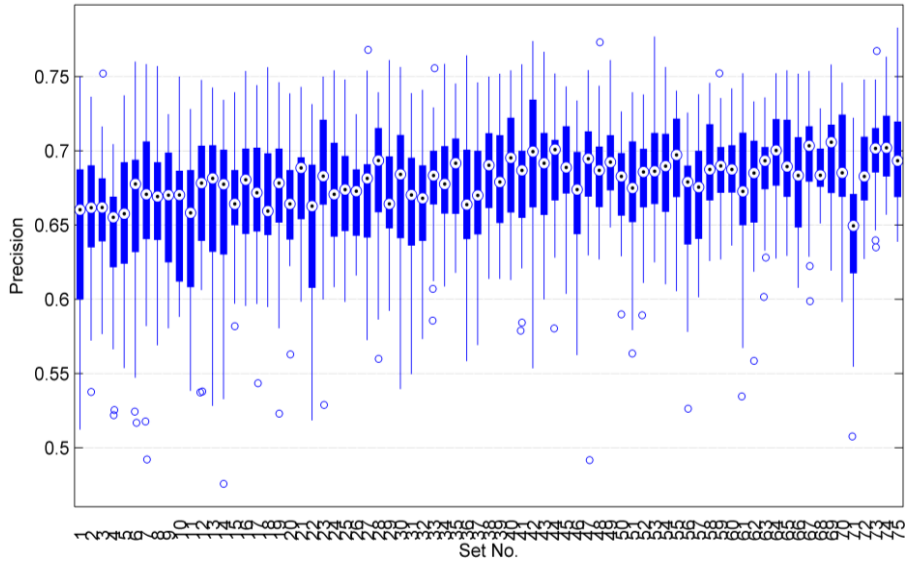
Fig. 10. Box plot of the precision of the proposed prediction model

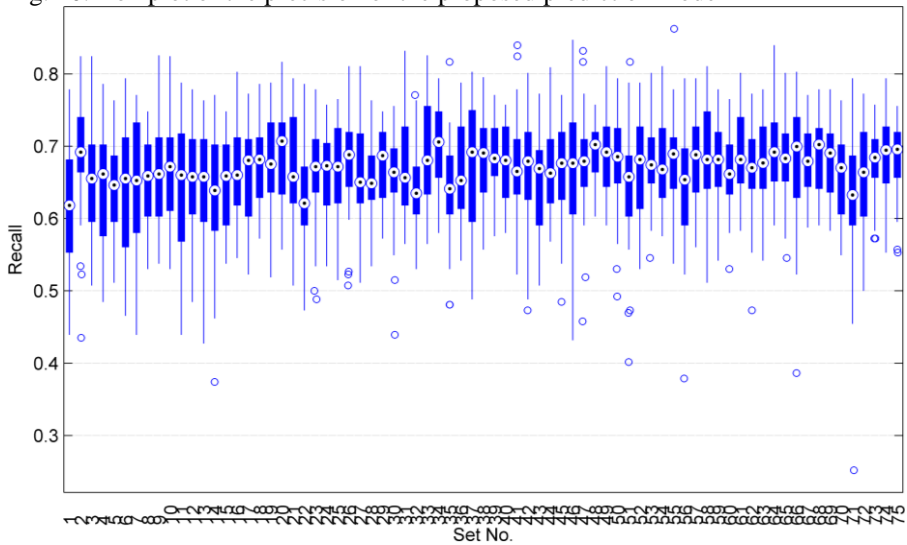
Fig. 11. Box plot of the recall of the proposed prediction model

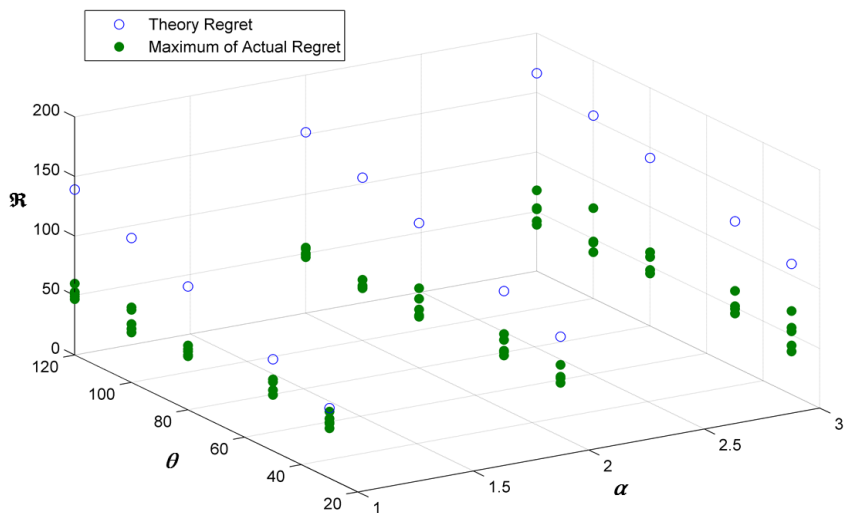
Fig 121. Illustration of theory regret and maximum of actual regret

From the results in Table 10, we can know that set numbers 53, 58, 67, and 74 provide better accuracy, precision, recall and elapsed time compared to the others.

Fig 12 illustrates the comparison between theory regret and maximum of actual regret for 75 sets. For the sets which have value of $\theta$ is smaller than 80, and value of $\alpha$ is 1, the gap between our theory regret and maximum value of actual regret is not so large. However, as the values of $\theta$ and $\alpha$ increase, the theory regret overestimates the maximum value of actual regret.

As mentioned above, we compared the performance of our proposed prediction model (OLR) with two well-known learning algorithms SVM and traditional logistic regression (TLR). The comparison is based on average value of mean and standard deviation of Accuracy, Precision and Recall. For SVM and TLR, we used RapidMiner to conduct the experiments with cross-validation. For OLR, we chose the result from experiment of set number 74.

Table 11. Comparison of SVM, TLR and ORL

| Method | Accuracy | Precision | Recall |
|---|---|---|---|
| SVM | 0.69±0.16 | 0.71±0.00 | 0.56±0.26 |
| TLR | 0.74±0.16 | 0.72±0.00 | 0.74±0.27 |
| OLR | 0.67±0.03 | 0.70±0.03 | 0.69±0.06 |

The results from Table 11 show that TLR outperforms SVM in Accuracy and Recall measurement. For Precision they are likely to be similar. Thus, logistic regression based method is more suitable for the proposed measurements set. Even though the OLR performance is somewhat inferior to SVM and TLR, it is acceptable to us considering the fact that OLR is online learning algorithm. Since SVM and TLR are batch-learning algorithms, they perform well with training set of labeled data. However, their performances for future instances of labeled data cannot be guaranteed. By contrast, our OLR can update its parameter vectors in order to deal with future instances. This is an indication of the flexibility and adaptability of the OLR algorithm.

**5. Conclusions and future work**

This paper presented a methodology that can be used in crowdsourcing applications for mining potential ideas. The proposed methodology consists of three main contributions. First, an extended text mining method that combines with computational linguistics techniques is used to extract information from crowdsourcing applications. Second, a set of measurements is used in order to transform text information into numerical value. Finally, we can use these measurements to predict the probability of being potential of an idea using the proposed online logistic regression. A case study of Dell's IdeaStorm was examined and numerical experiments were conducted to demonstrate the performance of the proposed methodology.

This methodology provides two main advantages to the companies. First, using the proposed text mining method and measurements, the companies can reduce the time for screening a ton of ideas from crowdsourcing application. Based on the absorptive capacity of the companies, they can combine the proposed text mining method with its rules and knowledge and hence enhancing the quality of result. Second, the proposed online logistic regression algorithm can play as a recommendation agent in the crowdsourcing. Each time a decision is made, the algorithm can learn its new parameter vectors. Over the time, if the decision behavior is stable, the set of parameter vectors can be convergent. Thus, when a new idea comes up, it can predict the probability of being potential of this idea. Different from traditional logistic regression algorithm, the proposed algorithm only needs a single training sample instead of a batch training sample. As time varies, the learning parameters may be changed to adapt the new decision.

For further researches, larger amount of data from the crowdsourcing applications should be used in order to have a better understanding about the real world behavior of the crowd as well as the decision maker's. The set of measurements should be refined to reflect this fact. Next, the proposed text mining method needs more rules in order to extract useful information including the relevance score more accurately. Combining text mining with sentiment analysis to analyze the comments is another further issue on text mining in crowd applications. This is because some comments may give additional useful information to the ideas, the relevance score may be varied by time if we consider sentimental states of users.

The proposed online mode logistic regression also needs to be improved for better quality. When data are examined, we only know the beginning state of the idea and the end state of the idea. However, if we have access to the full data and see how the states of the idea change (how the relevance and interest measurements are changed) and how decision makers response to this kind of "movement". We can utilize a prediction method in a better way.

**6. Acknowledgments**

This work was supported by a National Research Foundation of Korea grant (No. 2012R1A2008335) funded by the Korean Government. The fourth author was supported by the Basic Science Research Program through the National Research Foundation of Korea (NRF) funded by the Ministry of Education, Science and Technology (2010-0025650).

# Appendix

**Proof of Theorem 1**

Suppose the algorithm has already made totally ($m$-1) mistakes when the ($i-1$)-th instance arrives. Let the total weight value at ($i-1$)-th instance be:

$$H_{i-1} = w_1 + w_2 + \cdots + w^* + \cdots + w_{|H_{i-1}|}, \text{ where } w^* = \min_{j \leq |H_{i-1}|}\{w_j\}$$

Now, suppose our prediction on ($i-1$)-th instance makes $m$-th mistake, and that our algorithm samples the parameter vector which is associated with $w^*$ to use for its prediction. Hence, $H_{i-1}$ must be updated to a new value $H_i^*$, such that:

$$H_i^* = w_1 + w_2 + \cdots + w^* e^{-\alpha} + \cdots + w_{|H_{i-1}|} + 1$$

We prove that $H_i^*$ is the upper bound of total weight value at the $i$-th instance. Without loss of generality, we consider the case that our algorithm samples the parameter vector which is associated with $w_1$, thus:

$$H_i = w_1 e^{-\alpha} + w_2 + \cdots + w^* + \cdots + w_{|H_{i-1}|} + 1$$

Now, we get:

$$H_i^* - H_i = (1 - e^{-\alpha})(w_1 - w^*) \geq 0 \Rightarrow H_i^* \geq H_i$$

The above inequality holds for all $w_j, j \leq |H_{i-1}|$.

In addition, we have $w^* \geq e^{-\alpha(m-1)}$. If $w^* = e^{-\alpha(m-1)}$, $H_{i-1}$ becomes $H_{i-1} = e^{-\alpha(m-1)} + m - 1$. Hence, we get $H_i^* = e^{-\alpha m} + m \Rightarrow H_i \leq e^{-\alpha m} + m$ \hfill (*)

We split $D$ into two sub-sequences $D_1$ and $D_2$:

For the sequence $D_1 = \{(x_1, y_1), (x_2, y_2), \ldots, (x_{i^*-1}, y_{i^*-1})\}$, let $M_1$ be the total number of mistakes made by our algorithm on $D_1$ and let $m_1$ be the minimum number of mistakes made by any parameter vector of $B$ on the same sequence. Then, the following regret $\Re_1$ holds:

$$\Re_1 = M_1 - m_1 \leq \theta - 1$$

For the sequence $D_2 = \{(x_{i^*}, y_{i^*}), (x_{i^*+1}, y_{i^*+1}), \ldots, (x_n, y_n)\}$, we have $B = \{\beta_1, \beta_2, \ldots, \beta_\theta\}$. Suppose the total weight value at any the $i$-th instance, $i^* \leq i \leq n$ is as follows:

$$H_i = w_1 + w_2 + \ldots + w_\theta$$

Define function $I_{ij}$ to return 1 if at $i$-th instance the algorithm samples $\beta_j$ for its prediction, and such prediction incurs a loss value which satisfies the condition in line 8 of Algorithm 3. Otherwise, it returns 0. Considering the update of total weight value at the ($i+1$)-th instance:

$$H_{i+1} = w_1 e^{-\alpha I_{i1}} + w_2 e^{-\alpha I_{i2}} + \cdots + w_\theta e^{-\alpha I_{i\theta}}$$

Let $H_{i+1}/H_i$, then we have:

$$\frac{H_{i+1}}{H_i} = \frac{w_1}{H_i} e^{-\alpha I_{i1}} + \frac{w_2}{H_i} e^{-\alpha I_{i2}} + \cdots + \frac{w_\theta}{H_i} e^{-\alpha I_{i\theta}}$$

From the above equation, the amount $w_j/H_i$ can be seen as the probability of occurrence of $e^{-\alpha I_{ij}}$. Therefore, we rewrite:

$$\frac{H_{i+1}}{H_i} = \sum_{j=1}^{\theta} p_j e^{-\alpha I_{ij}} = E\left[e^{-\alpha I_i}\right]$$

Where $p_j = \frac{w_j}{H_i}$ and $E[\ ]$ are the probability of occurrence and the expectation of $e^{-\alpha I_{ij}}$, respectively.

$I_i$ indicates whether the prediction of our algorithm at the $i$-th instance makes a mistake. Hence, $I_i$ is a binary variable which has value 0 or 1. Now, we can apply Hoeffding's inequality [65]:

$$\ln \frac{H_{i+1}}{H_i} = \ln E\left[e^{-\alpha I_i}\right] \leq -\alpha E[I_i] + \frac{\alpha^2}{8}$$

Thus

$$\sum_{i=i^*}^{n} \ln \frac{H_{i+1}}{H_i} \leq -\alpha \sum_{i=i^*}^{n} E[I_i] + \frac{(n-i^*+1)\alpha^2}{8}$$

$$\ln\left(\frac{H_{i^*+1}}{H_{i^*}} \times \frac{H_{i^*+2}}{H_{i^*+1}} \times \cdots \times \frac{H_{n+1}}{H_n}\right) \leq -\alpha \sum_{i=i^*}^{n} E[I_i] + \frac{(n-i^*+1)\alpha^2}{8}$$

$$\ln \frac{H_{n+1}}{H_{i^*}} = \ln H_{n+1} - \ln H_{i^*} \leq -\alpha \sum_{i=i^*}^{n} E[I_i] + \frac{(n-i^*+1)\alpha^2}{8}$$

$$\ln H_{n+1} \leq -\alpha \sum_{i=i^*}^{n} E[I_i] + \frac{(n-i^*+1)\alpha^2}{8} + \ln H_{i^*}$$

Let $M_2 = \sum_{i=i^*}^{n} E[I_i]$ be the total number of mistakes our algorithm makes on $D_2$., and $m_2$ be the minimum number of mistakes made by any parameter vector of $B$ on the same sequence. Next, we get $\ln H_{n+1} \geq \ln\left(e^{-\alpha m_2}\right) = -\alpha m_2$ and $\ln H_{i^*} \leq \ln\left(e^{-\alpha(\theta-1)} + \theta - 1\right)$ (Equation (*)), and $i^* \geq \theta$. Hence, the following regret $\Re_2$ holds:

$$-\alpha m_2 \leq -\alpha M_2 + \frac{(n-i^*+1)\alpha^2}{8} + \ln\left(e^{-\alpha(\theta-1)} + \theta - 1\right)$$

$$\Re_2 = M_2 - m_2 \leq \frac{(n-\theta+1)\alpha}{8} + \frac{\ln\left(e^{-\alpha(\theta-1)} + \theta - 1\right)}{\alpha}$$

Therefore, the total regret $\Re$ is:

$$\Re = \Re_1 + \Re_2 \leq \theta - 1 + \frac{\alpha(n-\theta+1)}{8} + \frac{\ln\left(e^{-\alpha(\theta-1)} + \theta - 1\right)}{\alpha} \quad \blacksquare$$